# Glow-to-Arc Transition in Graphite cathode with High-Current Magnetron Discharge

Erwan Morel[1,2], Yoann Rozier[2], Tiberiu Minea[1]

[1] Université Paris Saclay, CNRS, Laboratoire de Physique des Gaz et des Plasmas, (LPGP), - Orsay, France
[2] Supergrid Institute, 23 Rue Cyprian, 69100 Villeurbanne, France

E-mail : erwanmorel@hotmail.fr; tiberiu.minea@universite-paris-saclay.fr

Abstract:

*The glow-to-arc transition is a critical phenomenon in plasma discharges, commonly leading to detrimental effects. The physical mechanisms triggering this transition remain poorly understood. The advent of a new discharge called Hyper Power Impulse Magnetron has opened new possibilities. HyPIM allows the glow mode to be maintained over long periods (1 ms) and at high current densities (>5 $A.cm^{-2}$), has unveiled certain features in the glow-to-arc transition. This work focuses on a graphite target that transits easily in the arc regime. The high-speed video-camera analysis revealed specific properties of graphite in ExB discharges, and the statistical study of the arc transition revealed differences from other refractory target materials. The early stage of cathodic spot formation, observed as bright dots, will be presented and analyzed within the known "ecton" and "vaporization" models for spot formation. This experimental study highlights the role of luminous spot formation prior to arc transition, with possible optimization on the stability of magnetron discharges.*

It is widely admitted that the diversity of experimental conditions leading to the abrupt change of a glow discharge into an electric arc involves several intricate mechanisms **[1],** as recently reviewed by Anders **[2]**. For instance, it has been demonstrated that thermionic emission from the cathode of a discharge can initiate a continuous transition. This emission can result from direct cathode heating due to positive ion bombardment or by raising the cathode temperature with an auxiliary heater. However, numerous experiments have shown that a transition can occur so quickly that all significant heating effects are excluded. Furthermore, this mechanism does not apply to a low melting point cathode. Several potential alternative mechanisms have been proposed **[3]**. Many of these involve extraneous effects, such as loose particles or various impurities on the cathode, like oxides or field effect emissions due to the spike effect associated with roughness.

Adding a magnetic field pushed back the glow-to-arc transition regarding current density **[4]**. Among the discharges confined by a magnetic field, so-called **ExB** discharges, the High-Power Impulse Magnetron Sputtering (HIPIMS) is particularly interesting because it operates at high current density. This discharge has been developed over the last 20 years as part of the ionized physical vapor deposition (I-PVD) technologies **[5].** Indeed **[6]**, HiPIMS uses a pulsed power supply delivering a huge peak power for a limited time (often ~10 µs) that exceeds the time-averaged power by typically two to three orders of magnitude but keeping for the same average power on the target as in conventional magnetron. This implies using long off-time periods separating two consecutive pulses (low-duty cycle) **[7]**. The peak power density, averaged over the target area, often exceeds 10⁷ W/m², aiming at the ionization of the sputtered atoms. Both neutral and ionized sputtered species leave the target as neutrals, and some get ionized in the plasma.

A typical HiPIMS operation uses a negative pulsed voltage applied to the cathode (or target) while the anode is grounded. Due to the magnetic topology in front of the cathode, the electrons follow cyclotron trajectories around the magnetic field lines combined with an azimuthal **ExB** drift oriented orthogonal to the plane formed by **B** and **E** [8]. This trajectory effectively confines the plasma electrons, augmenting their lifetime in the plasma. As a result, the gas ionization rate and sputtering efficiency both increase.

HiPIMS discharge with a graphite target has been studied earlier with argon as the process gas [9-13] and pure Ne or Ar/Ne mixture atmosphere [14-16]. They reported on the behavior of the discharge current, the recycling gas in glow mode, and the final properties of deposited thin films. Previous studies have shown that magnetron discharges in the glow regime often and easily transit to the arc. Even the origin of HiPIMS to limit the pulse duration to tens of microseconds is to avoid the glow-to-arc transition, keeping the discharge in an abnormal glow regime (high-voltage, high-current). People carrying out thin-film deposition seek to control this transition by controlling the current density [17].

Morel *et al*. recently demonstrated the possibility of reaching very high current levels but preserving the glow regime for long periods of time (1 ms) in a mode called Hyper-Power Impulse Magnetron (HyPIM) [18]. This mode is similar to HiPIMS but allows current densities to be 6 times higher and can only be reached under specific conditions (helium gas; pre-ionization (150 mA); limited pulse voltage (~320 V)). These conditions have been applied here to a graphite target. HyPIM with graphite leads to high current densities ($5 - 7$ $A.cm^{-2}$). However, increasing the current beyond this limit forces the arc transition for C targets, as revealed by different diagnostics.

This letter aims to investigate the transition from glow to cathodic arc. The involved mechanisms are qualitatively discussed, leading to low-voltage, high-current discharge formation with a graphite target. The typical current and voltage pulse waveforms have been analyzed in light of the plasma evolution recorded with a high-speed gated video camera in correlation with the magnetic field.

Our configuration is based on a standard unbalanced magnetron cathode (GENCOA VT50) hosted in a stainless-steel chamber (35 cm diameter and 28 cm height). Before discharge ignition, the chamber was evacuated to a base pressure <$2.10^{-6}$ Pa by a water-cooled turbomolecular pump (Edwards EXT 501/ISO 160) combined with a dual-stage rotary pump (Pfeiffer DUO 10M). The base pressure was measured by a full-range active Pirani compact cold cathode vacuum gauge (Pfeiffer PKR 361), and the pressure of helium (99.9995 % in purity) used as plasma gas was measured by a Baratron® capacitance manometer (MKS 627 BX). A throttle valve was installed between the vacuum chamber and the turbomolecular pump to adjust the pressure to the desired level.

Square voltage pulses were delivered by an Ionautics pulsing unit (HiPSTER 10) fed by an Elektro-Automatik DC power supply (PS 9000 3U), up to -1 kV. A Digital Delay Generator controlled pulse frequency and duration (Stanford Research system, model DG645). Voltage and current signals were recorded with an oscilloscope (Teledyne Lecroy, model WaveRunner 64Xi, 600 MHz bandwidth of 5 $Gs.s^{-1}$ sampling frequency ). A high-voltage probe acquired the voltage signal, while the current probe, incorporated into the pulsing unit, gave the pulse current. The pre-ionization direct current (DC) source was managed with a Technix negative power supply (SR1KV-300W, 1.5 KV, and 200 mA). Prior to the square pulse voltage, a DC voltage supported a constant low current discharge of 150 mA, called pre-ionization. Pulses superposition is managed through a box that contains two sets of diodes connected in parallel (Ionautics, 5 A diode box). With this electrical scheme, when one diode lets the current pass, the other blocks, and vice versa, enable a pulsed voltage to be superimposed on a DC voltage. The time evolution of the discharge glow was recorded in the full spectrum light with a high-speed camera (Phantom, model VEO710L). The current density is normalized to the whole target surface (50.8 mm diameter, 3 mm thick). The experimental setup is presented in **Fig. 1**.

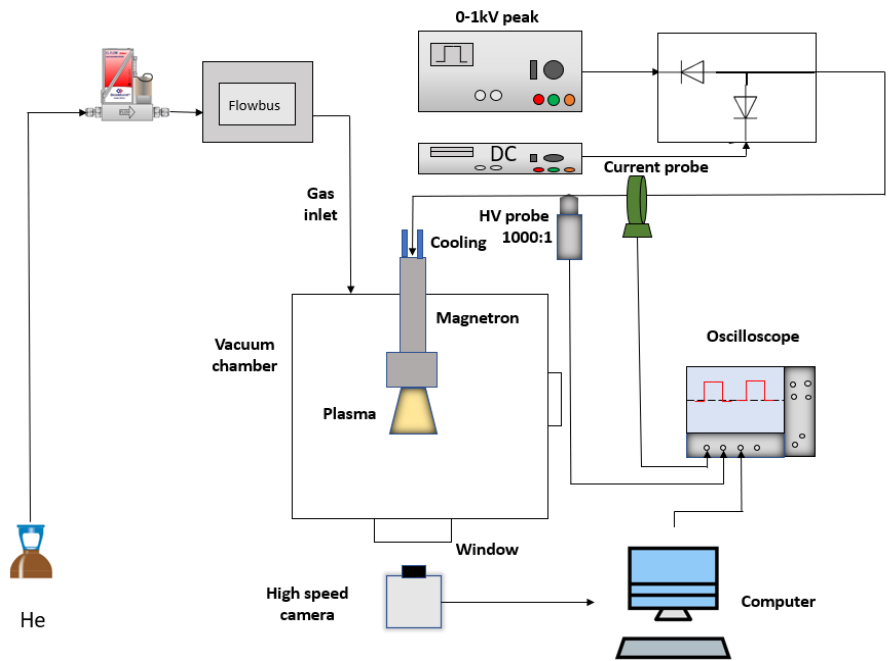

*Fig. 1: Sketch of experimental set-up and diagnostics arrangement.*

Statistical measurements were carried out with a graphite target to quantify the arc transition. Trains of 1,000 pulses (10 series of 100 pulses) were followed at a rate of 1 Hz. It was determined how many times the discharge evolved to arc. For comparison, the same tests were performed with a molybdenum target. To eliminate any transitory phenomena linked to oxides or roughness, the targets were previously cleaned with an argon plasma, and it should be recalled that the HyPIM mode uses a helium pre-ionization system to stabilize the discharge.

The results are shown in **Fig. 2.(a)**. Despite the very high current density of $31\ \text{A}\cdot\text{cm}^{-2}$ obtained with Mo, less than 10% of HyPIM pulses passed to the arc, whereas at a lower current density of $7\ \text{A}\cdot\text{cm}^{-2}$, most pulses (91.2%) with graphite experienced an arc. It is clear here that the arc transition is not due to the current density but other properties involving the cathode material. In our experimental configuration, for a given gas, the only way to modify the current density (without changing the magnetic configuration or the pressure, of course) is to adjust the pulse voltage. It is interesting to note that for graphite when the voltage is reduced by 20 V, the current density decreases by $2\ \text{A}\cdot\text{cm}^{-2}$, but the probability of arcing is divided by 9. One key parameter of estimating whether a material transits easily through the arc is its cohesive energy.

Cohesion energy represents the binding energy of atoms composing the solid. Providing this amount of energy to the target can release one atom in the vapor phase. The higher this energy, the more difficult the 'breaking' of the surface bonds is, limiting the surface damage. Therefore, the switch to the arc is delayed for materials with higher cohesion energy, so it is the cathodic spot appearance. In this sense, both cohesive energy and sublimation energy are very close. The experimental values of the cohesive energies ($E_{ce}$) are given by Anders **[19]**, and the theoretical sublimation energy ($E_s$) writes as equation (1). They are plotted in **Fig. 2.(b)**.

$$E_s = C_{th} \times T_{boil} + H_{vap} \tag{1}$$

where $C_{th}$ is the isobaric heat capacity, $T_{boil}$ is the boiling temperature, and $H_{vap}$ is the vaporization enthalpy.

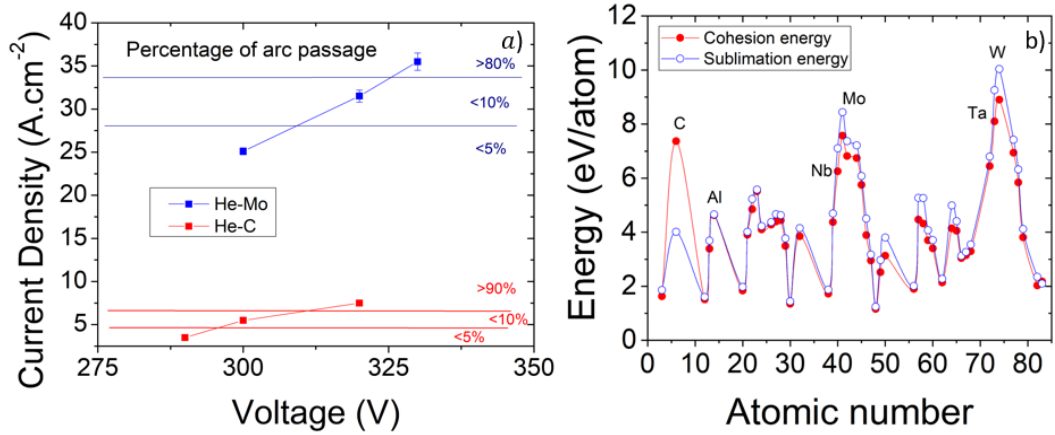

Fig. 2 : Statistical result of arc transition for graphite and molybdenum (a) and energy of cohesion and sublimation versus the atomic number (b).

What is striking in **Fig.2.(b)** is that three materials dominate the cohesive energy: carbon (C), molybdenum (Mo), and tungsten (W). Molybdenum and tungsten are two refractory metals with melting temperatures of 2632°C and 3422°C, respectively. Conversely, graphite has no melting point and goes directly from solid to gaseous if the temperature exceeds 3622 °C. So, theoretically, these three materials should be the best to avoid the transition to arc mode. Nevertheless, as in other works, it has been previously demonstrated that molybdenum and tungsten allow passing much higher currents than graphite without switching to arc **[20]**. So, despite a higher sublimation temperature, carbon arc transition is facilitated compared to refractory metals.

It is interesting to note in **Fig. 2.(*b*)** that the sublimation energy is slightly higher than the cohesive energy in the case of molybdenum and tungsten. In contrast, for carbon, it is twice lower. Consequently, the graphite sublimates for half of the energy required to break the bonds between the atoms. This means the material can sublimate before a spot formation, creating micro-protrusions that favor the arc transition. To fully understand why a high-current carbon discharge easily transits to an arc, it is essential to observe its dynamics closely.

Hence, images were taken for two discharges with comparable current density, one at 300 V (8.8 % probability to switch to arc, 5 $A.cm^{-2}$) and the other at 320 V (91.2% probability, 7 $A.cm^{-2}$) shown in **Fig. 3.(A)**. In the specific case of a voltage impulse of 320 V, one could observe, thanks to a more favorable transition probability, either a glow mode or an arc mode. These images reveal two pieces of information. First, during a high current pulse maintained in glow mode (Pictures **1** and **2, Fig. 3)**, numerous pre-spots or 'luminous dots' (as called hereafter) randomly cover the cathode surface. These points emitting light are at rest, immovable despite the cathode magnetic field.

Furthermore, for the 300 V pulse, these luminous dots occupy a very reduced part of the racetrack (distributed between the colored circles). In comparison, for the 320 V pulse, they cover a larger racetrack surface, and their brightness increases.

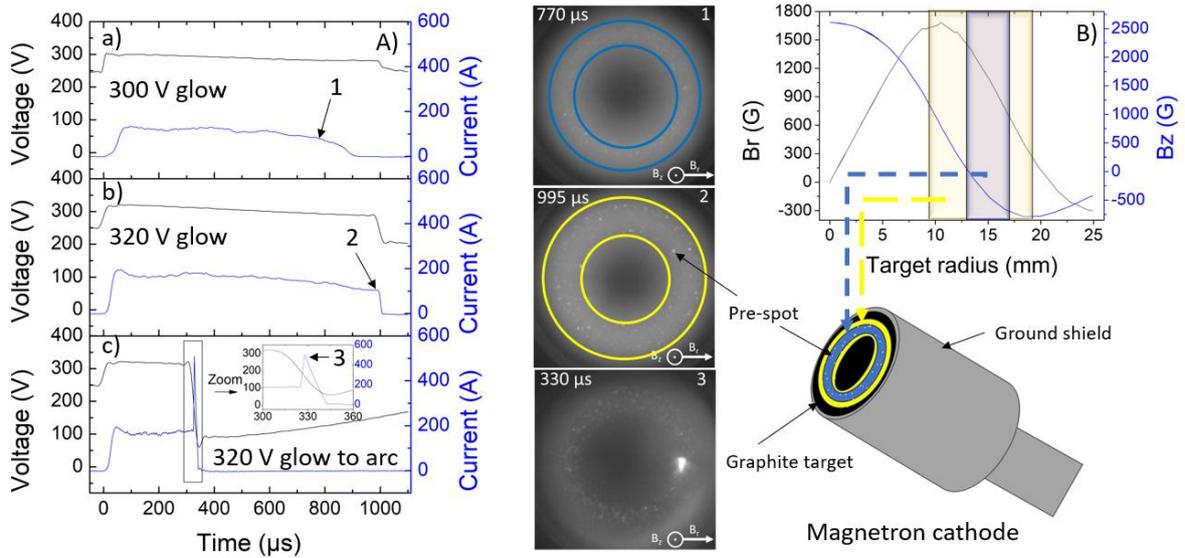

*Fig. 3 : Panel (A) the current and voltage waveforms during a typical pulse for different voltages in glow mode at 300 V (a) and 320 V (b,c). Pictures 1 and 2 show the glow mode for 300V and 320 V, respectively, and picture 3 shows the arc mode at 320 V. Panel (B) depicts the radial ($B_r$) and transverse ($B_z$) magnetic field at the target surface.*

When the discharge evolves in an arc regime, a very bright light emission area appears, degenerating in a cathode spot (Picture **3, Fig. 3 [3]**). The electrical characteristics for both glow mode and arc mode have been monitored in **Fig. 3**. The previously shown screenshot is marked on the curves with arrows. The voltage and current waveforms are relatively smooth throughout the pulse, except during the transition to arc. The immediate consequence of this transition is a rapid drop in discharge voltage, leading to plasma extinction within 5 µs. This occurs because the power supply is not designed to operate under arc conditions, and its protection disconnects the power. Despite this, the video camera is fast enough (200k FPS -Frame Per Second), taking an image every 5 µs. Hence, the birth of the cathodic spot can be analyzed before the discharge extinguishes. One clearly observed the cathode spot starting to move under the effect of the magnetic field. This phenomenon is well known and is called Robson's deviation, and the cathodic spot moves in the direction of **JxB [21].**

Two theories have been proposed to explain how a cathodic spot works: the "vaporization" model and the "ecton" model **[22]**. The first assumes electrons to be emitted under the combined action of a strong electric field and a high metal temperature at the spot. This emission is accompanied by the evaporation of neutral atoms from the cathode, further ionized by the energetic electrons. Most of the ions thus produced feed the inter-electrode plasma, and only a fraction of them return to the cathode, their impact on the cathode surface contributing to the local rise in the cathode temperature **[23]**.

The "ecton" model, or explosive electron emission model, postulates that the cathode spot is made up of a set of very small cells with a very short lifetime (~10 ns) **[24]**. Under the effect of an explosive-type process involving a small volume of metal in the cathode, these cells directly produce a plasma made up of electrons and ionized (metal) vapor particles. These two theories are still debated in the literature.

The fast video camera reveals that the luminous dots present before the arc transition would support the "ecton" model of the spot creation. Still, in the HyPIM plasma, the explosive aspect of a spot takes place over several hundred µs, unlike the "ecton" model where the phenomena occur over a few nanoseconds. According to the "ecton" model **[25-27]**, thermal runaway originates on micro protrusions until they explode with a delay time associated with the material-specific action. From the theory of wire explosions **[28]**, the current density, *J*, and explosion delay time, *dt*, satisfy:

$$\int_0^{td} J^2 dt = \bar{h} \qquad (2)$$

where $\bar{h}$ is called the specific action, whose value depends on the cathode material but is approximately independent of current density (*J*), wire cross-section, or other discharge quantities.

The specific action value for carbon is $1.8 \; A^2.s.m^{-4}$. If one uses equation (2) with the current value in **Fig. 3. (A.c)** and assumes that the current is uniformly distributed over the whole target, the lifetime of the luminous dots until one of them could transit into the cathode spot (in the recorded movie of the discharge pulse: 325-110=215µs), then the numerical calculation of the integral gives $1.46 \times 10^6 \; A^2.s.m^{-4}$. Clearly, the "ecton" model alone cannot explain the observed dots-to-spot transition phenomena.

According to Anders **[29]**, the two mechanisms of "vaporization" and "ecton" could follow one another. An initial explosive emission phase could be followed by a second, longer phase, during which metal evaporation maintained by the power source comes into play. Moreover, the evaporation of the metal would be favored by the local rise in temperature of the metal in the vicinity of the spot following the explosive process. In our case, it seems to be the other way around: the luminous dots, which are not cathodic spotlights, are still hot because they emit light. They vaporize the target until one explodes, leading to a cathode spot. This is possible because the sublimation energy is much lower than the cohesive energy in the case of carbon (**Fig. 2.(*b*))**. A modeling study, considering the physics of field emission at micro and nanometric emitters, showed that graphite behaved differently from a refractory material (tungsten). Although it is difficult to compare this study with the experimental results presented here, it was clear that graphite and refractory materials have very different electron emission and arc transition behavior despite very similar melting and sublimation temperatures **[30].**

Moreover, graphite is a specific material. The observed light spots may be because graphite is a second-class conductor with lower conductivity than metals. In addition, the target under test is polycrystalline (conductivity drops at the grain boundaries), with many facets and misalignments affecting the target's mechanical and thermal cohesion. This means that certain points on the surface can be insulating (or highly resistive), giving rise to the local deposition of electrical charges, or can rise in temperature due to the nature of the surface bonds which lead to these luminous dots

They are all potential ignition centers (i.e., surfaces that can pass to a cathodic spotlight), increasing the probability of an arc transition and the appearance of one or more dominant cathode spots. Tracking them with dedicated software accounts for the number of light dots. They appear uniformly and randomly distributed along the **ExB** racetrack area after 100 µs and are observed throughout the discharge. Moreover, the cathode luminous spots extinguish 10 µs after the end of the pulse. In the case of the 300 V discharge with 8.8 % arc transition probability, 175±30 light dots were identified in the blue area on the cathode represented in **Fig. 3**. In comparison, the number of dots often exceeds 400 ± 50 in the case of the 320 V discharge with 91.2 % arc transition probability (yellow area on the cathode in **Fig. 3**). Thus, more light dots strongly correlate with an increased likelihood of arc transition and the higher the current density, the greater the **ExB** racetrack area occupied by these luminous dots.

Another phenomenon that can affect the arc transition is the topology of the magnetic field. **Fig. 3.(B)** shows the area occupied by the luminous dots with the intensity of the magnetic field along the target radius, in blue (**picture 1**) for the 300 V discharge and in yellow (**picture 2**) for the 320 V discharge. It is well known in the literature **[4]** that the component parallel to the target ($B_r$) plays a significant role in magnetron plasma confinement and arc transition. Strengthening $B_r$ decreases the arc transition probability and increases the discharge current while remaining in glow mode. Here, the magnetic field doesn't seem to impact the arc transition or the spread of the luminous dots. The $B_r$ field dominates in the yellow area, with the strongest arc transition. The significant aspect of the transition to arc mode with a graphite target is essentially due to the presence of luminous dots mentioned above and the number of these pre-spots during discharge.

In conclusion, this work provides new insights into the easier transition in arc mode for graphite magnetron cathodes. Statistical studies and comparisons with refractory metals such as tungsten and molybdenum have shown that the arc transition is not directly related to the current density but rather to the material properties. The specific discharge used (HyPIM) with long pulse time (1 ms) and high current densities (>5 $A.cm^{-2}$) have revealed new phenomena. For graphite, sublimation energy is half of its cohesion energy, allowing the creation of numerous pre-spots or luminous dots. This phenomenon, revealed by the images collected with a high-speed video camera, explains why graphite transits easily in the arc regime. These dots of light cover more and more the racetrack and, therefore, increase the probability of an arc transition and the appearance of one or more dominant cathode spots after a few hundred μs. Unlike cathodic spots, these pre-spots are not sensitive to magnetic fields and stay stationary. Beyond the graphite specificities, this study advances the understanding of the glow-to-arc transition, identifying pre-spots as a critical intermediate stage. The observed pre-spots can be modeled using fractals **[31]**, a mathematical method to model electric arcs. This new phenomenon will help to deepen the understanding of glow-to-arc transitions.

### Acknowledgment

This work was supported by a grant from the French National Research Agency (ANR) as part of the "Investissement d'Avenir" Program (ANE-ITE-002-01).

### Data Availability

The data supporting this study's findings are available by contacting the corresponding author upon reasonable request.